\documentstyle[12pt]{article}
\begin{document}
\newcommand{\de}{\delta}\newcommand{\ga}{\gamma}
\newcommand{\e}{\epsilon} \newcommand{\ot}{\otimes}
\newcommand{\be}{\begin{equation}} \newcommand{\ee}{\end{equation}}
\newcommand{\ba}{\begin{array}} \newcommand{\ea}{\end{array}}
\newcommand{\beq}{\begin{equation}}\newcommand{\eeq}{\end{equation}}
\newcommand{\tmod}{{\cal T}}\newcommand{\amod}{{\cal A}}
\newcommand{\bemod}{{\cal B}}\newcommand{\cmod}{{\cal C}}
\newcommand{\dmod}{{\cal D}}\newcommand{\hmod}{{\cal H}}
\newcommand{\s}{\scriptstyle}\newcommand{\tr}{{\rm tr}}
\newcommand{\einsop}{{\bf 1}}
\def\oR{R^*} \def\upa{\uparrow}
\def\R{\overline{R}} \def\doa{\downarrow}
\def\dag{\dagger}
\def\ve{\epsilon}
\def\si{\sigma}
\def\ga{\gamma}
\newcommand{\reff}[1]{eq.~(\ref{#1})}
\centerline{\bf{Integrable generalised spin ladder models}}
~~~\\
\begin{center}
{\large Angela Foerster$^{1}$, Jon Links$^{1,2}$ and Arlei Prestes
Tonel$^{1}$}
~~~\\
{$ \phantom{0000000000}$}

{\em ${}^1$Instituto de F\'{\i}sica da UFRGS \\ Av. Bento
Gon\c{c}alves
9500,
Porto Alegre, RS - Brazil }
~~~\\
{$ \phantom{0000000000}$}

{\em ${}^2$Department of Mathematics \\
University of
Queensland, Queensland, 4072,  Australia}
\end{center}
~~~\\
\begin{abstract}
We present two new integrable spin ladder models which posses three
general free parameters besides the rung coupling $J$.
Wang's systems based on the $SU(4)$ and $SU(3|1)$
symmetries can be obtained as special cases. The models are
exactly solvable by means of the Bethe ansatz method.
\end{abstract}
%
%
\vfil\eject

Recently there has been a great interest on spin ladder systems
from both theoretical and experimental point of view for their
relevance to some quasi-one dimensional materials, which under
hole doping may exhibit superconductivity \cite{dagotto}.
These systems are reasonably well approximated by Heisenberg
ladders, which take into account only couplings along
the legs and the rungs. 
Although these systems are not exactly solvable, a variety
of solvable ladder models have been found \cite{rod}, \cite{albeverio},
\cite{batchelor1}.
Of particular interest is a general 2-leg spin ladder
system with biquadratic interactions proposed by Wang \cite{wang},
which by suitable choices of the interchain and interrung
coupling originates two integrable spin ladder models based
on the $SU(4)$ and $SU(3/1)$ symmetries. In these cases
the rung interactions appear as chemical potentials which
break the underlying symmetries of the models.
Subsequently other generalised integrable spin ladders
have been proposed in the literature 
\cite{batchelor,frahm,jon,kolezhuk,fei,angi}.
In these cases, no (or few) free parameters are present due
to the strict conditions of integrability.

The purpose of this paper is to present two new integrable
generalized spin ladders with three extra
parameters without violating integrability. These
models are exactly solvable by the Bethe ansatz and they
reduce to Wang's models \cite{wang}
for a special limit of these extra parameters.

Let us begin by introducing the first generalised spin ladder model,
whose Hamiltonian reads
\be
H^{\s{{(1)}}}=\sum_{j=1}^{L} \biggl[ \, h_{j,j+1} +
{\s{\frac{1}{2}}} J
\left( \vec{\sigma_{j}}.\vec{\tau_{j}}-1 \right) \,
\biggr]
\label{ha}
\ee
where
\begin{eqnarray}
&h_{j,j+1}&=\sigma_{j}^{+}\sigma_{j+1}^{-} \,
\biggl[ \, \frac{t_{1}^{-2}}{4} (1+\tau_{j}^{z})(1+\tau_{j+1}^{z})
+ \frac{t_{2}^{2}}{4}(1-\tau_{j}^{z})(1-\tau_{j+1}^{z})
+ t_{3}^{2}\, \tau_{j}^{+}\tau_{j+1}^{-}+\tau_{j}^{-}\tau_{j+1}^{+} \,
\biggr] \quad \nonumber \\ & & +
\sigma_{j}^{-}\sigma_{j+1}^{+} \,
\biggl[\, \frac{t_{1}^{2}}{4}(1+ \tau_{j}^{z})(1+\tau_{j+1}^{z})
+ \frac{t_{2}^{-2}}{4}(1-\tau_{j}^{z})(1-\tau_{j+1}^{z})
+ \tau_{j}^{+}\tau_{j+1}^{-} + t_{3}^{-2}\, \tau_{j}^{-}\tau_{j+1}^{+}
\,
\biggr] \quad \nonumber \\ & &+
{\s{\frac{1}{4}}}(1+\sigma_{j}^{z})(1+\sigma_{j+1}^{z}) \,
\biggl[\, {\s{\frac{1}{2}}}(1+\tau_{j}^{z}\tau_{j+1}^{z})
+t_{1}^{-2}\, \tau_{j}^{+}\tau_{j+1}^{-}+t_{1}^{2} \,
\tau_{j}^{-}\tau_{j+1}^{+} \,
\biggr] \quad \nonumber \\ & &+
{\s{\frac{1}{4}}}(1-\sigma_{j}^{z})(1-\sigma_{j+1}^{z}) \,
\biggl[\, {\s{\frac{1}{2}}}(1+\tau_{j}^{z}\tau_{j+1}^{z})
+t_{2}^{2}\, \tau_{j}^{+}\tau_{j+1}^{-}+ \, t_{2}^{-2} \tau_{j}^{-}
\tau_{j+1}^{+} \,
\biggr].  \quad \nonumber
\end{eqnarray}
Above $\vec{\sigma_{j}}$ and $\vec{\tau_{j}}$ are Pauli matrices acting
on site $j$ of the upper and lower legs, respectively, $J$ is the
strength of the rung coupling and $t_{1},t_{2}, t_{3}$ are
general independent parameters. $L$ is the number of rungs and
periodic boundary conditions are imposed. By setting $t_{1}, t_{2},
t_{3} \rightarrow 1$ in equation (\ref{ha}), Wang's model based
on the $SU(4)$ symmetry \footnote{strictly speaking, it is SU(4) in the 
absence of the rung coupling}\cite{wang} can be recovered.

The integrability of this model can be shown by the fact that it can
be mapped to the Hamiltonian below,
which can be derived from an $R-$matrix obeying the
Yang-Baxter algebra for $J = 0$, while for
$J\neq 0$ the rung interactions take the form of a chemical
potential term.
\begin{equation}
{\hat{H}}^{{\s(1)}} = \sum_{j=1}^{L}
\biggl[ {\hat h}_{j,j+1} - 2J \, X^{00}_{j} \biggr]
\label{ha1}
\end{equation}
where
\begin{eqnarray}
&{\hat h}_{j,j+1}&=\sum_{\alpha =0}^{3}X^{\alpha 
\alpha}_{j}X^{\alpha \alpha}_{j+1}+
X^{2 0}_{j}X^{0 2}_{j+1}+X^{0 2}_{j}X^{2 0}_{j+1} \quad \nonumber \\ & & 
+
t_{1}^{2}\biggl(X^{1 0}_{j}X^{0 1}_{j+1}+
X^{1 2}_{j}X^{2 1}_{j+1}\biggr)+t_{2}^{2}\biggl(X^{3 0}_{j}
X^{0 3}_{j+1}+X^{3 2}_{j}X^{2 3}_{j+1}\biggr)+t_{3}^{2}
X_{j}^{3 1}X_{j+1}^{1 3}  \quad \nonumber \\ & & +
t_{1}^{-2}\biggl(X^{0 1}_{j}X^{1 0}_{j+1}+
X^{2 1}_{j}X^{1 2}_{j+1}\biggr)+t_{2}^{-2}\biggl(X^{0 3}_{j}
X^{3 0}_{j+1}+X^{2 3}_{j}X^{3 2}_{j+1}\biggr)+t_{3}^{-2}
X_{j}^{1 3}X_{j+1}^{3 1}  .
\nonumber
\end{eqnarray}
Above $ X^{\alpha \beta}_{j} = |\alpha_{j}><\beta_{j}|$
are the Hubbard operators with $ |\alpha_{j}>$ the orthogonalised
eigenstates of the local operator $\vec{\sigma_{j}}.\vec{\tau_{j}}$,
as in Wang's case \cite{wang}.

The following $R$-matrix
\begin{equation}
\label{r}
{\footnotesize
R=\pmatrix{a&0&0&0&|& 0&0&0&0&|& 0&0&0&0&|& 0&0&0&0&\cr
               0&t_{1}^{-2}b&0&0&|& c&0&0&0&|& 0&0&0&0&|& 0&0&0&0&\cr
               0&0&b&0&|& 0&0&0&0&|& c&0&0&0&|& 0&0&0&0&\cr
               0&0&0&t_{2}^{-2}b&|& 0&0&0&0&|& 0&0&0&0&|& c&0&0&0&\cr
               -&-&-&-& & -&-&-&-& & -&-&-&-& & -&-&-&-&\cr
               0&c&0&0&|& t_{1}^{2}b&0&0&0&|& 0&0&0&0&|& 0&0&0&0&\cr
               0&0&0&0&|& 0&a&0&0&|& 0&0&0&0&|& 0&0&0&0&\cr
               0&0&0&0&|& 0&0&t_{1}^{2}b&0&|& 0&c&0&0&|& 0&0&0&0&\cr
               0&0&0&0&|& 0&0&0&t_{3}^{-2}b&|& 0&0&0&0&|& 0&c&0&0&\cr
               -&-&-&-& & -&-&-&-& & -&-&-&-& & -&-&-&-&\cr
               0&0&c&0&|& 0&0&0&0&|& b&0&0&0&|& 0&0&0&0&\cr
               0&0&0&0&|& 0&0&c&0&|& 0&t_{1}^{-2}b&0&0&|& 0&0&0&0&\cr
               0&0&0&0&|& 0&0&0&0&|& 0&0&a&0&|& 0&0&0&0&\cr
               0&0&0&0&|& 0&0&0&0&|& 0&0&0&t_{2}^{-2}b&|& 0&0&c&0&\cr
               -&-&-&-& & -&-&-&-& & -&-&-&-& & -&-&-&-&\cr
               0&0&0&c&|& 0&0&0&0&|& 0&0&0&0&|& t_{2}^{2}b&0&0&0&\cr
               0&0&0&0&|& 0&0&0&c&|& 0&0&0&0&|& 0&t_{3}^{2}b&0&0&\cr
               0&0&0&0&|& 0&0&0&0&|& 0&0&0&c&|& 0&0&t_{2}^{2}b&0&\cr
               0&0&0&0&|& 0&0&0&0&|& 0&0&0&0&|& 0&0&0&a&\cr} } \, \, \,
,
\end{equation}
with
$$a=x+1\, \, \, \, ,b=x,\, \, \, \,  c=1,$$
obeys the Yang-Baxter algebra
\begin{equation}
R_{12}(x-y)R_{13}(x)R_{23}(y)=R_{23}(y)R_{13}(x)R_{12}(x-y)
\end{equation}
and originates the Hamiltonian (\ref{ha1}) for $J=0$ by the
standard procedure
$$ {\hat h}_{j,j+1}= P \frac{d}{dx}R(x)|_{x=0}, $$
where $P$ is the permutation operator.

The model can be solved exactly by the Bethe ansatz
method and the Bethe ansatz equations read
\begin{eqnarray}
t_{1}^{2(L-M_{3})}t_{2}^{2M_{3}}t_{3}^{-2M_{3}}\left(\frac{\lambda_{j}-i/
2}
{\lambda_{j}+i/2}\right)^{L}&=&\prod_{l\neq
j}^{M_{1}}\frac{\lambda_{j}-\lambda_{l}-i}
{\lambda_{j}-\lambda_{l}+i}
\prod_{\alpha=1}^{M_{2}}\frac{\lambda_{j}-\mu_{\alpha}+i/2}
{\lambda_{j}-\mu_{\alpha}-i/2} \nonumber \\
t_{1}^{2(L-M_{3})}t_{2}^{2M_{3}}t_{3}^{-2M_{3}}\prod_{\beta\neq 
\alpha}^{M_{2}}\frac{\mu_{\alpha}-\mu_{\beta}-i}
{\mu_{\alpha}-\mu_{\beta}+i}&=&\prod_{j=1}^{M_{1}}\frac{\mu_{\alpha}-\lambda_{j}-i/2}
{\mu_{\alpha}-\lambda_{j}+i/2}
\prod_{\delta =1}^{M_{3}}\frac{\mu_{\alpha}-\nu_{\delta}-i/2}
{\mu_{\alpha}-\nu_{\delta}+i/2}
\label{bae} \\
t_{1}^{2(M_{2}-M_{1})}t_{2}^{-2(L-M_{1}+M_{2})}t_{3}^{-2(M_{1}-M_{2}}\prod_{\gamma \neq
\delta}^{M_{3}}
\frac{\nu_{\delta}-\nu_{\gamma}-i}{\nu_{\delta}-\nu_{\gamma}+i} &=&
\prod_{\alpha =1}^{M_{2}}\frac{\nu_{\delta}-\mu_{\alpha}-i/2}
{\nu_{\delta}-\mu_{\alpha}+i/2}
\nonumber
\end{eqnarray}
 

The energy eigenvalues of the Hamiltonian (\ref{ha1}) are
given by
\begin{equation}
E=-\sum_{j=1}^{M_{1}}\biggl(\frac{1}{\lambda_{j}^{2}+1/4}-2J\biggr)
+\frac{3}{4}\left(1-2J\right)L
\end{equation}
where $\lambda_{j}$ are solutions to the Bethe ansatz equations
(\ref{bae}).


Now let us introduce a second integrable spin ladder model with three 
extra parameters, whose Hamiltonian reads
\be
{\cal{H}}=\sum_{j=1}^{L}
\biggl[k_{j,j+1}-\frac{1}{4}(1-2J)(\vec{\sigma_{j}}.\vec{\tau_{j}}-1)\biggr]
\label{ha2}
\ee
where
\begin{equation}
k_{j,j+1}=h_{j,j+1}-\frac{1}{8}(\vec{\sigma_{j}}.\vec{\tau_{j}})(\vec{\sigma_{j+1}}.\vec{\tau_{j+1}})
+\frac{1}{4}\left(\vec{\sigma_{j}}.\vec{\tau_{j}}-1\right) \,\,\,\,
\end{equation}
and $h_{j,j+1}$ is given by eq. (1).

The solvability of Hamiltonian above lies in the fact that it can
be mapped, as before, to a Hamiltonian which can be derived for
an $R$-matrix satisfying the Yang-Baxter algebra for $J=0$, while
for $J \neq 0$ the rung interactions take the form of a chemical
potential term
\begin{eqnarray}
{\hat {\cal{H}}}=\sum_{j=1}^{L}\biggl[{\hat k}_{j,j+1} 
+(1-2J)X^{00}_{j}\biggr]
\label{ha21}
\end{eqnarray}
where
\begin{eqnarray}
&{\hat k}_{j,j+1}&=\sum_{\alpha =0}^{3}X^{\alpha 
\alpha}_{j}X^{\alpha \alpha}_{j+1}
-2\, X^{0 0}_{j}X^{0 0}_{j+1}+X^{2 0}_{j}X^{0 2}_{j+1}+X^{0 2}_{j}X^{2 
0}_{j+1} \quad \nonumber \\ & & +
t_{1}^{2}\biggl(X^{1 0}_{j}X^{0 1}_{j+1}+
X^{1 2}_{j}X^{2 1}_{j+1}\biggr)+t_{2}^{2}\biggl(X^{3 0}_{j}
X^{0 3}_{j+1}+X^{3 2}_{j}X^{2 3}_{j+1}\biggr)+t_{3}^{2}
X_{j}^{3 1}X_{j+1}^{1 3}  \quad \nonumber \\ & & +
t_{1}^{-2}\biggl(X^{0 1}_{j}X^{1 0}_{j+1}+
X^{2 1}_{j}X^{1 2}_{j+1}\biggr)+t_{2}^{-2}\biggl(X^{0 3}_{j}
X^{3 0}_{j+1}+X^{2 3}_{j}X^{3 2}_{j+1}\biggr)+t_{3}^{-2}
X_{j}^{1 3}X_{j+1}^{3 1}  .
\end{eqnarray}

For $J=0$ the model is derived by standard methods from the following
${\cal{R}}$-matrix
\begin{equation}
\label{r}
{\footnotesize
{\cal{R}}=\pmatrix{w&0&0&0&|& 0&0&0&0&|& 0&0&0&0&|& 0&0&0&0&\cr
               0&t_{1}^{-2}b&0&0&|& c&0&0&0&|& 0&0&0&0&|& 0&0&0&0&\cr
               0&0&b&0&|& 0&0&0&0&|& c&0&0&0&|& 0&0&0&0&\cr
               0&0&0&t_{2}^{-2}b&|& 0&0&0&0&|& 0&0&0&0&|& c&0&0&0&\cr
               -&-&-&-& & -&-&-&-& & -&-&-&-& & -&-&-&-&\cr
               0&c&0&0&|& t_{1}^{2}b&0&0&0&|& 0&0&0&0&|& 0&0&0&0&\cr
               0&0&0&0&|& 0&a&0&0&|& 0&0&0&0&|& 0&0&0&0&\cr
               0&0&0&0&|& 0&0&t_{1}^{2}b&0&|& 0&c&0&0&|& 0&0&0&0&\cr
               0&0&0&0&|& 0&0&0&t_{3}^{-2}b&|& 0&0&0&0&|& 0&c&0&0&\cr
               -&-&-&-& & -&-&-&-& & -&-&-&-& & -&-&-&-&\cr
               0&0&c&0&|& 0&0&0&0&|& b&0&0&0&|& 0&0&0&0&\cr
               0&0&0&0&|& 0&0&c&0&|& 0&t_{1}^{-2}b&0&0&|& 0&0&0&0&\cr
               0&0&0&0&|& 0&0&0&0&|& 0&0&a&0&|& 0&0&0&0&\cr
               0&0&0&0&|& 0&0&0&0&|& 0&0&0&t_{2}^{-2}b&|& 0&0&c&0&\cr
               -&-&-&-& & -&-&-&-& & -&-&-&-& & -&-&-&-&\cr
               0&0&0&c&|& 0&0&0&0&|& 0&0&0&0&|& t_{2}^{2}b&0&0&0&\cr
               0&0&0&0&|& 0&0&0&c&|& 0&0&0&0&|& 0&t_{3}^{2}b&0&0&\cr
               0&0&0&0&|& 0&0&0&0&|& 0&0&0&c&|& 0&0&t_{2}^{2}b&0&\cr
               0&0&0&0&|& 0&0&0&0&|& 0&0&0&0&|& 0&0&0&a&\cr} } \, \, \,
,
\end{equation}
with
$$a=x+1\, \, \, \, ,b=x,\, \, \, \,  c=1\, \, \, \,  , w=-x+1,$$
which obeys the Yang-Baxter algebra.
The above Hamiltonian  has a similar algebraic structure as that of an
SU(3/1) supersymmetric t-J model. Using the algebraic nested Bethe 
ansatz
method this model can be solved and
the Bethe ansatz equation are given by
\begin{eqnarray}
t_{1}^{2(L-M_{3})}t_{2}^{2M_{3}}t_{3}^{-2M_{3}}\left(\frac{\lambda_{j}-i/
2}
{\lambda_{j}+i/2}\right)^{L}&=&\prod_{\alpha
=1}^{M_{2}}\frac{\lambda_{j}-\mu_{\alpha}-i/2}
{\lambda_{j}-\mu_{\alpha}+i/2} \nonumber \\
t_{1}^{2(L-M_{3})}t_{2}^{2M_{3}}t_{3}^{-2M_{3}}\prod_{\beta\neq 
\alpha}^{M_{2}}\frac{\mu_{\alpha}-\mu_{\beta}-i}
{\mu_{\alpha}-\mu_{\beta}+i}&=&\prod_{j=1}^{M_{1}}\frac{\mu_{\alpha}-
\lambda_{j}-i/2}{\mu_{\alpha}-\lambda_{j}+i/2}
\prod_{\delta =1}^{M_{3}}\frac{\mu_{\alpha}-\nu_{\delta}-i/2}
{\mu_{\alpha}-\nu_{\delta}+i/2} \\
t_{1}^{2(M_{2}-M_{1})}t_{2}^{-2(L-M_{1}+M_{2})}t_{3}^{-2(M_{1}-M_{2})}\prod_{\gamma \neq
\delta}^{M_{3}}
\frac{\nu_{\delta}-\nu_{\gamma}-i}{\nu_{\delta}-\nu_{\gamma}+i}&=&
\prod_{\alpha =1}^{M_{2}}\frac{\nu_{\delta}-\mu_{\alpha}-i/2}
{\nu_{\delta}-\mu_{\alpha}+i/2} \nonumber \\
\end{eqnarray}

The eigenenergy of the Hamiltonian  (9) is given by

\begin{equation}
E=\sum_{j=1}^{M_{1}}\biggl (\frac{1}{\lambda_{j}^{2}+1/4}+2J-1 
\biggr)-2JL
\end{equation}
above $\lambda_{j}$ are solutions of Bethe-ansatz equations (12).


To summarize, we have introduced a new generalization of Wang's spin 
ladder models
based on the $SU(4)$ and $SU(3|1)$ symmetries. This was achieved
by introducing three extra parameters into the system without
violating integrability. The Bethe ansatz equations as well as
the energy expressions of the models were presented.
The physics of the integrable models presented here is expected to
be of interest, since the presence of these extra parameters
may turn the phase diagram of the models much richer.

\centerline{{\bf Acknowledgements}}
~~\\

JL thanks the Funda\c{c}\~{a}o de Amparo a Pesquisa do Estado do Rio
Grande do Sul and Australian Research Council for financial support. He also
thanks the
Instituto de F\'{\i}sica da UFRGS for their kind hospitality.
AF and APT thank CNPq-Conselho Nacional de Desenvolvimento
Cient\'{\i}fico e
Tecnol\'ogico for financial support.



\begin{thebibliography}{99}
\bibitem{dagotto} E. Dagotto and T.M.Rice, Science {\bf 271}, 618 (1996)
\bibitem{rod} H. Frahm and C. R\"odenbeck, Europhys. Letters {\bf 33}, 
47 (1996)
\bibitem{albeverio}S. Albeverio, S.-M. Fei and Y.Wang, Europhys.
Lett. {\bf 47} 364 (1999)
\bibitem{batchelor1}M.T.Batchelor and M. Maslen, Ground state energy and 
mass gap of a generalized quantum spin ladder, cond-mat/9907480
\bibitem{wang} Y. Wang, Phys. Rev. B {\bf 60}, 9236 (1999)
\bibitem{batchelor}M.T.Batchelor and M. Maslen, J.Phys. A: Math.Gen. 
{\bf 32} (1999) L377.
\bibitem{frahm}H. Frahm and A. Kundu, Phase diagram of an exactly 
solvable
t-J ladder model, cond-mat/9910104.
\bibitem{jon} M.T.Batchelor, J. de Gier, J. Links and M. Maslen,
cond-mat/9911043
\bibitem{kolezhuk} A.K. Kolezhuk and H.-J. Mikeska, Int. J. Mod. Phys. 
{\bf B12}
(1998) 2325
\bibitem{fei} S. Albeverio and S.M. Fei, Exactly solvable models
of generalized spin-ladders, cond-mat/9807341.
\bibitem{angi} J. Links and A. Foerster, Solution of a two leg
spin ladder system, cond-mat/9911096, Phys. Rev. B, to appear
\end{thebibliography}
\end{document}